# In-plane Antiferromagnetism in Ferromagnetic Kagome Semimetal $Co_3Sn_2S_2$

Sandy Adhitia Ekahana[1], Satoshi Okamoto[2], Jan Dreiser[3], Loïc Roduit[5], Gawryluk Dariusz Jakub[5], Andrew Hunter[4], Anna Tamai[4], Y. Soh[6*]

1 Quantum Photon Science, Paul Scherrer Institute, Forschungstrasse 111, CH-5232, Villigen, Switzerland

2 Materials Science and Technology Division, Oak Ridge National Laboratory, Oak Ridge, Tennessee 37831, USA

3 Swiss Light Source, Paul Scherrer Institute, Forschungstrasse 111, CH-5232, Villigen, Switzerland

4 Department of Quantum Matter Physics, University of Geneva, 24 Quai Ernest-Ansermet, CH-1211, Geneva, Switzerland

5 Laboratory for Multiscale Materials Experiments (LMX), Paul Scherrer Institute, Forschungstrasse 111, CH-5232, Villigen, Switzerland

6 Paul Scherrer Institute, Forschungstrasse 111, CH-5232, Villigen, Switzerland

*To whom correspondence should be addressed. Email: yona.soh@psi.ch

**Abstract:**

$Co_3Sn_2S_2$ has been reported to be a Weyl semimetal with broken time-reversal symmetry with c axis ferromagnetism (FM) below a Curie temperature of 177 K. Despite the large interest in $Co_3Sn_2S_2$, the magnetic structure is still under debate and recent studies have challenged our understanding of the magnetic phase diagram of $Co_3Sn_2S_2$ by reporting unusual magnetic phases including the presence of exchange bias. Understanding the magnetism of $Co_3Sn_2S_2$ is important since its electronic band structure including the much-celebrated flat bands and Weyl nodes depend on the magnetic phase. In this work, using X-ray Magnetic Circular Dichroism (XMCD), we establish that the magnetic moment in Co arises from the spin, with negligible orbital moment. In addition, we detect an in-plane AFM minority phase in the sea of a FM phase using spatially-resolved angle-resolved photoemission spectroscopy ($\mu$-ARPES) combined with density functional theory (DFT) calculation. Separately, we detect a sharp flat band precisely at the Fermi level ($E_F$) at some regions in the sample, which we attribute to a surface state. The AFM phase survives even to the low temperature of 6 K. This example of entirely different magnetic ground states in a stoichiometric intermetallic invites further efforts to explore the observed AFM phase and understand the origin and nature of the magnetic and electronic inhomogeneity on the mesoscale and the interface between the AFM and FM phases.

**Introduction**

Magnetic frustration frequently manifests in materials with a kagome lattice structure where the inherent shared triangle configuration poses a challenge to satisfy certain magnetic interactions to minimize the total energy, giving rise to an AFM fluctuation [1], spin fluctuation [2], and other frustrated magnet situations [3-7]. Meanwhile, a kagome FM $Fe_3Sn_2$ with a unidirectional magnetic moment that undergoes spin reorientation from the c-axis towards the kagome plane at a lower temperature [8] settles comfortably at a certain minimum energy state without suffering magnetic frustration. $Co_3Sn_2S_2$, which is also reported to be a c-axis FM [9-11] with $T_C \cong 177$ K and a saturated magnetic moment of 0.3 $\mu_B$ per Co [10], almost has that privilege if not due to several reports that contradict the simple FM phase paradigm.

There are conflicting reports regarding the magnetic phase diagram of $Co_3Sn_2S_2$ although the searches of Weyl points were conducted [12-16] assuming such a FM phase. The sample in the first report of the Weyl points [17] shows a hysteresis loop typical for a FM (square and symmetric shape, no exchange bias) at low temperature with no report of an unusual magnetic phase. However, it was reported in 2017 that $Co_3Sn_2S_2$ harbors an anomalous magnetic phase just below the Curie temperature ($T_c \sim 175$ K) and above a temperature called $T_A \approx 125$ K (or in the temperature range 125 – 175 K) before it adopts a fully FM phase with magnetic moments along the *c*-axis as long as the applied external magnetic field is small [18]. Increasing this external magnetic field makes this anomalous phase disappear and shows a pure FM phase.

This unique phase was revisited through muon spin rotation ($\mu$SR) experiments and reported to be a competition between the FM phase and an in-plane AFM phase, whose phase diagram depends on the doping concentration of indium in $Co_3Sn_{2-x}In_xS_2$, with increasing In promoting a mixed phase of FM and AFM [19, 20]. On the other hand, neutron scattering experiments on $Co_3Sn_{2-x}In_xS_2$ concluded that instead of a phase coexistence of pure FM and AFM phases, the system is a *homogeneous* phase starting from a pure c axis FM evolving to a canted FM with the canting angle from the c axis increasing from 0° to 65° as the indium doping is varied from 0 to 30% [21, 22]. Further DFT calculations suggest that the cobalt magnetic moment may be canted by a small angle of 1.5 ° off the c-axis at 128 K [23] whereas a study based on pair distribution functions on powder suggests a local instability distortion (random canting) of the Co moment to be ∼ 20 ° off the perpendicular axis [24] while maintaining an average out-of-plane moment in the long range order. Recent studies with non-linear optics report a homogeneous but temperature dependent magnetic phase, with a c axis FM at $T_A < T < T_c$ and a canted FM below $T_A$ [25]. Therefore, it is still a debated question whether an AFM phase coexists with the FM phase at $T_A < T < T_c$ and whether the magnetic configuration in $Co_3Sn_2S_2$ at $T < T_A$ is a pure *c*-axis out-of-kagome-plane FM or a canted FM. Furthermore, scanning tunneling spectroscopy (STS) studies reported a negative orbital moment of $-3\ \mu_B$ arising from a flat band at $E_F$ in $Co_3Sn_2S_2$, which seems to be at odds with the small magnetic moment of 0.3 $\mu_B$ per Co.

We would like to stress the difficulty of detecting a minority AFM phase in the sea of a FM majority phase when the propagation vector, *k*, of both considered magnetic maximal symmetry Shubnikov supgroups is equal to [0, 0, 0]. Due to the "hidden" nature of the AFM order not exhibiting any net magnetic moment, when AFM is a small minority phase in the presence of an FM majority phase, techniques such as magnetomery fail since the signal is dominated by the FM phase. When an AFM order is characterized by a propagation vector other than [0, 0, 0], in addition to nuclear Bragg peaks associated with the crystal, new purely magnetic reflections appear in the magnetically ordered state enabling easier detection of the AFM order if the magnetic moment is sufficiently large. However, when k = [0, 0, 0] and the magnetic moment is small, which is the case of $Co_3Sn_2S_2$, detecting a small amount of a minority AFM phase is challenging since there are no new reflections and the locations of the AFM reflections coincide with those of the nuclear Bragg and FM peaks, which have much larger scattering intensities.

In this paper, we show the existence of AFM regions at $T = 6$ K by using micro-focus angle-resolved photoemission spectroscopy ($\mu$-ARPES) with 6.01 eV photon energy (4th harmonic continuous laser) in an otherwise majority FM phase. We also find out that the orbital moment in Co is negligible based on XMCD. We demonstrate that the observed AFM band structure is of magnetic origin by cycling the temperature up to the paramagnetic (PM) phase, where it disappears, and show that it reappears as we cool the system to $T = 6$ K. Our discovery opens a new perspective that the phase below $T_A$ is not entirely FM and a minority AFM phase in $Co_3Sn_2S_2$ exists even at a low temperature of 6 K. These findings enrich our understanding of AFM and FM coexistence as we spatially display the coexistence with direct visualization of the band structure owing to the advanced laser $\mu$-ARPES technique. Ultimately, this finding invites further exploration of the spin and band structure analysis of such topological kagome systems and opens paths for future engineering where the coexistence of different magnetic phases can be controlled and utilized in magnetic and electronic applications such as in exchange bias and spintronics.

**Crystal Structure and Magnetic Properties of $Co_3Sn_2S_2$**

The $Co_3Sn_2S_2$ compound crystallizes in Shandite-like structure of $Ni_3Pb_2S_2$ archetype [26] with space group R-3m, number 166, and lattice constant $a = 5.38$ Å and $c = 13.19$ Å in a conventional unit cell with hexagonal cross section. The composition of $Co_3Sn_2S_2$ in the conventional unit cell consists of three kagome layers translated from each other (figure 1a). Each kagome layer comprises of Co atoms that lie on the kagome lattice site with one Sn atom lying at the center of the Co kagome lattice or star of David shape. Meanwhile, each triangle forming the kagome lattice, which consists of Co atoms, hosts an S atom hovering above the triangle center of mass alternating on the top side (S1) and the bottom side (S2) (which have equivalent Wyckoff position (6c)) of the kagome lattice (green prisms in figure 1) as we circle around the triangles forming the star of David. On the same triangles, we have Sn atoms also hovering on the triangle center of mass projection but with longer bond lengths (blue prisms) than the S-kagome bond (green prisms) and placed opposite to the S atoms. The Sn atoms are shared between neighboring kagome layers on the top and on the bottom, making this $Co_3Sn_2S_2$ a quasi-layered structure due to these shared Sn atomic layers. From this structure, $Co_3Sn_2S_2$ allows two different cleaving planes: cleaving between the Sn-S layer or cleaving between the S-kagome layer. These two cleaving planes are not equivalent as the formation energy is smaller in breaking the Sn-S bond than the S-kagome bond [12, 16]. Therefore, it is rare to observe the kagome termination or the S termination whose layers underneath are the Sn layer and S layer consecutively (see SIFig.1. (iii)&(iv)) [16]. We can also deduce from a simple observation that the bond length between the Co atom on the kagome plane and the S atom on the nearby S1 or S2 layer is the shortest (2.20 Å) (green prism vertices connecting S and Co atoms). Therefore, it is more probable to have the S termination with the kagome layer underneath (S1) or the Sn termination.

For the magnetic structure, only the Co atoms give rise to the magnetic moment, where DFT calculations have concluded that the minimum energy configuration is an FM phase ($R\bar{3}m'$ magnetic space group No 166.101), where all Co moments are pointing out of the kagome plane (*c*-axis FM or $(0,0,m_z)$) [12, 27]. Our synchrotron-based ARPES result (with beam spot of ~50 $\mu m$ x 50 $\mu m$) in Fig. 1b also captures the FM bands, which agree with our DFT calculation ($\mu_{Co} = 0.35\ \mu_B$) if we introduce a "band renormalization" to simulate the correlation effect and raise the calculated band minimum closer to $E_F$ to match the ARPES data, as also performed in other publications [14, 28]. We do this by dividing the energy scale on the DFT result by a certain factor (1.52 in [28]) and thus raising the band energy position. Our synchrotron-based ARPES result agrees qualitatively with other published results [13, 29] showing the general consensus of ferromagnetism in this material. Realizing a pure AFM phase is not favorable energetically as both the "chiral" ($R\bar{3}m$ space group No 166.97) and the "in/out" magnetic configuration (also being realized in $R\bar{3}m'$ space group by two spin components of magnetic cobalt in *$m_x$,2$m_x$,$m_z$* configuration) (see Fig. 1a) require a constrained calculation [27]. Our DFT calculation shows that the FM phase is lower in energy by ~50 meV than both AFM phases, while the $R\bar{3}m$ AFM phase is reported to be lower in energy by ~1 meV than the $R\bar{3}m'$ AFM phase [27] ($E_{FM} < E_{AFM\ R-3m} < E_{AFM\ R-3m'}$)..

**X-ray Magnetic Circular Dichroism (XMCD)**

To explore the magnetic phase in Co$_3$Sn$_2$S$_2$ further, we measure the element specific magnetization at the cobalt $L_{2,3}$ edges with x-ray magnetic circular dichroism (XMCD) in the total electron yield (TEY) mode, performed at the X-Treme beam line at the Paul Scherrer Institute [30]. The advantage of studying the magnetic behavior using XMCD is that it can disentangle the orbital and spin magnetic moment contributions separately. The schematic of the XMCD experiment setup is given in Figure 2a showing that the photon $\vec{k}$-vector is coupled with the direction of the applied magnetic field (a positive magnetic field is opposite to the direction of the photon $\vec{k}$-vector and vice-versa), and the sample normal direction can be rotated for angle dependent measurements. With this measurement, we explore how the x-ray absorption spectrum (XAS) at the cobalt $L_{2,3}$-edges varies with left and right circularly polarized light (C- and C+) (figure 2a), from which we can calculate the XMCD signal from both $L_3$ and $L_2$ edges. The sample was freshly cleaved in-situ prior to the measurement at $T = 50$ K and at a pressure of $p \sim 10^{-11}$ mbar for the results presented in panel a,c, and d in figure 2. Meanwhile, the data shown in figure 2b was obtained on a carbon-capped sample after being freshly cleaved in another ultra-high vacuum (UHV) chamber and transferred in an atmospheric environment to the X-Treme chamber. This explains the quantitatively different sum-rule results we present below, where the carbon-capped sample shows a relatively smaller effective spin value than the freshly cleaved sample. However, the results from both samples agree qualitatively.

Our XAS result agrees with the published XANES data [31]. We can see in figure 2a that the XAS signals from C+ and C- show a visible difference, from which the XMCD = $I_{C+} - I_{C-}$ can be calculated; the integrated XMCD signal is proportional to the cobalt magnetization projected along the x-ray wavevector. We performed the sum-rule analysis (detailed explanation in SI) on the XMCD data to obtain the orbital ($m_l$) and effective spin magnetic moment ($m_{S_{eff}}$) as a function of temperature (figure 2b) at normal incidence and also as a function of the incidence angle (figure 2c) at $T = 50$ K. We can see in figure 2b that the orbital magnetic moment is relatively small at all temperatures while the effective spin moment vanishes at $T \sim 175$ K, in agreement with the magnetic transition towards the PM phase. The relatively negligible orbital moment implies that the magnetic moment in Co$_3$Sn$_2$S$_2$ comes mainly from the spin. This result is at odds with the reported large negative orbital magnetism ($\sim -3\mu_B$) inferred from the field dependence of the peak in the density of states (DOS) observed around $E_F$ by scanning tunneling spectroscopy (STS) [32]. We highlight here that XMCD is a well-established element-specific magnetic probe, and it provides a direct measurement of the orbital and effective spin moments, in contrast to the STS.

The angle-dependent result shows a decreasing value for the effective spin projected onto the x-ray direction at higher incidence angles $m_{Seff,60°} = \frac{2}{3} m_{Seff,0°}$ (figure 2c) showing a strongly anisotropic behavior agreeing with the preferred c-axis FM orientation. However, since $m_{Seff,60°}$ is not strictly $\frac{1}{2} m_{Seff,0°}$, it is an indication that either the spin is a little tilted from the surface normal, as a result of responding to the angled external field, and/or there is a contribution of the magnetic dipole term $\langle T_{\vec{k}} \rangle$, since $m_{S_{eff}} = -2\langle S_{\vec{k}} \rangle - 7\langle T_{\vec{k}} \rangle$ where $\langle S_{\vec{k}} \rangle$ is the expectation value of the electron spin projected along the $\vec{k}$ vector of the x-ray beam. Reference [33] discusses the $\langle S_{\vec{k}} \rangle$ and $\langle T_{\vec{k}} \rangle$ for a single kagome layer case and concludes that the pure AFM spin configuration with positive chirality of the spin (the in/out and chiral AFM condition in figure 1 without any canting to the c-axis) yields $\langle T_{\vec{k}} \rangle = 0$ [33]. Since the net $\langle S_{\vec{k}} \rangle$ is also 0 for the AFM configuration, we can assume that the XMCD signal we observe here is not from a pure AFM configuration. Meanwhile, the pure c-axis FM condition within the approximation of reference [33] will lead to a collinear $\langle S_{\vec{k}} \rangle$ and $\langle T_{\vec{k}} \rangle$, which are hard to be disentangled. The real situation in $Co_3Sn_2S_2$ can be more complex than what reference [33] describes.

We further confirm the strong magnetic anisotropy in $Co_3Sn_2S_2$ by showing the magnetic hysteresis loops in figure 2d measured at $T = 50$ K by following the XMCD at the $L_3$ peak. The applied magnetic field in figure 2b ($B = 1$ T) and figure 2c ($B = 6.8$ T) are large enough to saturate the magnetic moment in $Co_3Sn_2S_2$. The saturated XMCD signal (normalized to the L3 peak intensity) at a grazing incidence angle of 60° drops by $\frac{2}{3}$ compared to the 0° value agreeing with the sum-rule result. The sudden signal flip at the switching field suggests that the magnetization reversal in $Co_3Sn_2S_2$ occurs by domain wall motion and not by gradual rotation of the moments. Surprisingly, we observe an exchange bias, which typically occurs in an artificial system of layered FM and AFM structures [34], at both incident angles suggesting that $Co_3Sn_2S_2$ does not behave like a normal FM material. Instead, it shows the asymmetry of the magnetic moment along the c-axis direction (it needs a larger negative field strength to flip the signal than the positive field) even after saturation is seemingly reached, which suggests the presence of an AFM phase beside the FM phase at $T = 50$ K. . Such coexistence is unusual for a single crystal stoichiometric system. However, there has been a report on exchange bias in self-flux grown Co$_3$Sn$_2$S$_2$ which was attributed to a spin glass phase arising from geometric frustration of the kagome lattice [35]. A more recent report [36] on exchange bias in self-flux grown Co$_3$Sn$_2$S$_2$ speculates its origin to the existence of AFM at the domain walls.

**Laser ARPES**

The XMCD suggestion of an AFM phase demands further exploration of the system with spatial resolution capability. Up to now, standard spatially resolved magnetic probes such as magnetic force microscopy or magneto-optic Kerr effect techniques have not been successful in detecting AFM regions [37, 38]. For this, we utilize instead the spatial resolution of laser $\mu$-ARPES to investigate the fingerprint of the AFM phase at $T = 6$ K. In this work, we report three different cleaved surfaces that we call sample 1, sample 2, and sample 3 as shown in figure 3. It should be first noted that most of the area investigated from all three samples do not display sharp dispersing features and instead we observe a broad and smeared band. However, its integrated energy distribution curve (EDC) shows a signal with a typical shape as shown in figure 3 with a broad peak around $-0.225$ eV $< E_B < -0.175$ eV. In addition, we discover two independent features: a flat band right at $E_F$ and a rather broad band structure that we call the "butterfly" shape. They are independent of each other as each can be seen separately: flat band alone in sample 1, butterfly band alone in sample 3, and both observed together in sample 2. The areas where the flat band and/or the butterfly shape are observed are relatively rare for each cleaved plane (orange rectangle in figure 3). We have cleaved other samples whose ARPES signal is only the majority broad and smeared band, without any flat band at $E_F$ or the butterfly shape. We also note that in general, the synchrotron-based result shows no band like the butterfly within the momentum-energy window covered by the laser ARPES, in any $k_z$ position, as shown in Fig. 1b. Thus, the majority of the area shown by the laser ARPES agrees qualitatively with our synchrotron-based ARPES, which shows no visible band.

We proceed to analyze these findings with density functional theory (DFT) calculations and explore the origin of the butterfly shape and the flat band. At first, we notice that this butterfly shape is similar to the case of AFM R-3m and R-3m' calculation [27]. However, the energy of the bands does not match our ARPES data as the calculated band lies at a smaller binding energy (closer to $E_F$). The band energy renormalization that is discussed before for Fig. 1b and performed in [14, 28] will not resolve this issue since the bands need to move to a higher binding energy and not closer to $E_F$. We overcome this issue by introducing a larger magnetic moment used in the DFT calculations than $\mu_{Co} \approx 0.33\ \mu_B$. The value $\mu_{Co} \approx 0.33\ \mu_B$ has been used in several DFT calculations [12, 17, 27, 28] as it has been obtained from various bulk measurements [9, 18, 21, 32, 39]. However, this small value of cobalt magnetic moment raises a question about its microscopic origin as the cobalt atom magnetic moment can be as high as $\mu_{Co} \approx 2.2 - 2.7\ \mu_B$ for a system with a small orbital contribution [40, 41]. Meanwhile, in the case of Co-TiO$_2$, the reported cobalt moment is $\mu_{Co} \approx 0.32\ \mu_B$, which they attribute to the low spin state of the cobalt, *i.e.*, the magnetic moment is coming from the spin (not orbital) of the cobalt [42]. It has been reported that $\mu_{Co} = \frac{1}{3}\mu_B \approx 0.33\ \mu_B$ in the FM case is trivial from a shared cobalt triangle cluster [29]. However, the situation in the in-plane AFM case of Co$_3$Sn$_2$S$_2$ may lead to a different effective cobalt moment, which cannot be captured easily with the DFT.

Thus, we explore different DFT results as we vary the value of $\mu_{Co}$ and we settle with $\mu_{Co} \approx 0.6\ \mu_B$ as it matches the observed butterfly band shape and energy position (figure 3) that we probe at $k_z \sim 0.6\ \Gamma Z$ with 6.01 eV (see SI figures for details). Our ARPES data agrees better with the pure in-plane chiral AFM phase DFT result, where we only have a degenerate band at the $\bar{\Gamma}$ point (nodal line along the $k_z$ direction) as we can see only one broad peak at the EDC (figure 3). However, we cannot rule out the pure in/out AFM phase that predicts split bands at the $\bar{\Gamma}$ point, even though the splitting is not visible in the raw-ARPES result. Meanwhile, we associate the "blurred" band and yet a visible peak on the EDC in the majority area with the FM $\mu_{Co} \approx 0.35\ \mu_B$ calculation (and energy renormalization) by considering the incoherent peak scattered from the majority FM phase. Figure 3 sample 2 shows a clearer FM cut as compared to sample 1 and sample 3 majority regions, which demonstrate that this blurred band has some spatial variation and is also sample dependent.

We attribute the visible flat band near $E_F$ (figure 3) to a band in the chiral AFM phase originating from sulphur termination as suggested by our DFT calculation with $\mu_{Co} \approx 0.35\ \mu_B$ (see figure 3 and SIFig. 17 for more details). Difference in the effective magnetic moment for the surface and the majority bulk area ($\mu_{Co} \approx 0.35\ \mu_B$) vs the in-plane AFM bulk ($\mu_{Co} \approx 0.6\ \mu_B$) may seem controversial but it is plausible given the peculiar magnetic behavior of Co$_3$Sn$_2$S$_2$ and the precise nature of the magnetic ground state not being settled yet. Even for bulk, a moment of 0.39 $\mu_B$ rather than 0.33 $\mu_B$ was determined as the best fit for neutron scattering [21]. This flat band in the chiral AFM phase on sulphur termination coincides with the experimental finding by STS in a previous report [32]. However, in their case, they attributed the peak found around $E_F$, on both the S-termination (001) surface and the side surface but not on the Sn-termination (001) surface, to come from the bulk FM band at $k_z = 0$ (as also suggested by our DFT calculation in SIFig. 5). As this special flat band at the Fermi surface is only found at a particular area with ARPES and not anywhere else, it is possible that the flat band observed here is of surface rather than of bulk origin. It should be noted that this flat band near $E_F$ is not visible in our synchrotron based ARPES result further suggesting that this flat band is a rare occurrence. Whether this flat band found in ARPES is the same flat band reported previously [32] is open for discussion.

Finally, we demonstrate that the butterfly-shaped band is of magnetic origin by monitoring how the butterfly band at a fixed place (taken from sample 3 in figure 3) evolves as a function of temperature towards 200 K where it is known to have transitioned to the PM phase. We can see in figure 4 that the band structure gradually changes such that the peak of the EDC moves closer to $E_F$ while the band itself is still broad. This broad band energy position agrees with the PM phase energy position as suggested by the DFT result in the lower panel of figure 4, which is close to $E_F$. The origin of this broadness can be due to a trivial $k_z$ broadening as suggested by our slab calculation for the PM phase in SIFIg.17 while we do not exclude the possibility of magnetic disorder as suggested before [29]. Lastly, we further confirm the magnetic origin of the band by lowering the temperature back to the base temperature where we recover the butterfly shape located around the same area.

The combined findings from XMCD and $\mu$-ARPES suggest that the AFM phase exists even at low temperature in Co$_3$Sn$_2$S$_2$. While this AFM and FM coexistence at low temperature has been reported in doped compounds of Co$_3$Sn$_{2-x}$In$_x$S$_2$ with *x* ranging from 0.05 to 0.3 [20], this can be due to compositional/structural/strain inhomogeneity in a non-stoichiometric compound. Our result suggests existence of a minority AFM phase in the absence of doping in a stoichiometric compound. We also would like to point out that the published results about exchange bias in Co$_3$Sn$_2$S$_2$ [13, 35, 36] report it at low temperature as our ARPES measurement and even speculate how the AFM phase may form [35, 36], but our result is the first time such AFM phase is spatially mapped. Among the three different sample growth techniques which have been used, self-flux method [14, 21], Bridgeman technique [20, 22, 43, 44], and chemical vapor transport (CVT) technique (see SI for crystal growth technique), only two of them (self-flux method and CVT) have displayed exchange bias. It is beyond the scope of this paper to investigate the atomic origin that makes our CVT grown Co$_3$Sn$_2$S$_2$ to host both FM and AFM even at low temperature, while other CVT grown Co$_3$Sn$_2$S$_2$ [45] but with different technical steps exhibits a pure FM hysteresis loop. We rule out an impurity phase since none of the elements or binary compounds exhibit AFM. The phase coexistence of FM and AFM in Co$_3$Sn$_2$S$_2$ may have a similar origin to that observed in stoichiometric pyrochlore Yb$_2$Ti$_2$O$_7$[46], with both systems having the ingredients for magnetic frustration giving rise to multiphase competition. The absence of AFM phase in some of the other reported Co$_3$Sn$_2$S$_2$ studies may be due to the selection of the FM phase by disorder, *i.e.*, FM order by disorder [47].

**Conclusions and Future Outlooks**

In summary, we have demonstrated that AFM exists as a minority phase in Co$_3$Sn$_2$S$_2$ even at 6 K where Co$_3$Sn$_2$S$_2$ was previously assumed to be in a pure FM phase. In this work, the AFM phase is first indirectly suggested by the XMCD results showing an exchange bias in the hysteresis loop (a typical indicator of the presence of an AFM material next to a FM material) even at a temperature of 50 K, lower than the temperature range of 90 K - 175 K where AFM has been reported. Our spatially resolved laser $\mu$-ARPES in combination with the DFT calculation further confirms that there is an AFM phase co-existing as a minority phase with the majority FM phase even at 6 K. As ARPES only probes the first few layers of the compound, it is still an open question whether such pockets of pure AFM phase exist in the bulk of Co$_3$Sn$_2$S$_2$. This finding may play a role in the relatively small total saturated Co magnetic moment observed [10]. Our results still do not resolve the question of whether the AFM and FM phases compete at higher temperatures or doped compounds as reported by $\mu$SR or it is (mostly) a FM phase with the possibility of canting as suggested by neutron scattering and second harmonic generation experiments. Nevertheless, our work further enriches our understanding of the magnetism of Co$_3$Sn$_2$S$_2$, which is still controversial, and suggests an inhomogeneous electronic band structure associated to the coexistence of the AFM and FM phase, which is relevant in the context of Weyl semimetals and the interface of topologically different electronic domains. This work also invites further theoretical consideration of the spatial co-existence of two different phases in a single stoichiometric compound and further search for other materials that host such co-existence. Last, our finding that the orbital magnetic moment of Co in $Co_3Sn_2S_2$ is negligible should be taken into account when considering physical effects that rely on spin-orbit coupling.

**Acknowledgement**

S.A.E acknowledges the support from NCCR-MARVEL funded by the Swiss National Science Foundation, the European Union's Horizon 2020 research and innovation programme under the Marie Skłodowska-Curie grant agreement No 701647, and the European Research Council HERO Synergy grant SYG-18 810451. The research by S.O. was supported by the U.S. Department of Energy, Office of Science, Basic Energy Sciences, Materials Sciences and Engineering Division.

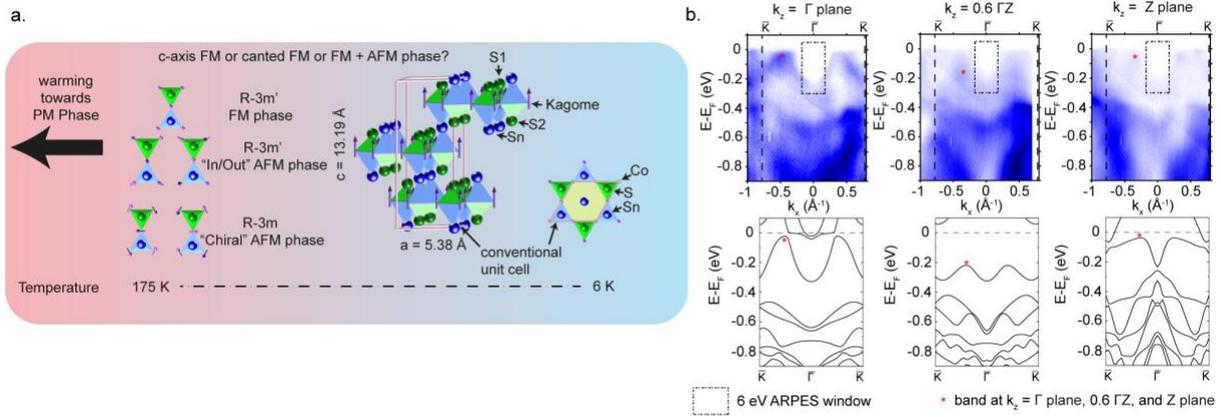

**Figure 1**. $Co_3Sn_2S_2$ crystal, magnetic structure, and $k_z$ dependent ARPES result. a. The crystal structure shows the kagome layers of Co-Sn with S and Sn atoms alternatingly occupying the center of the kagome triangle at a certain distance from the kagome layer creating a quasi-layer of S and a quasi-layer of Sn, which is shared between the kagome layers. The schematic shows the common understanding that $Co_3Sn_2S_2$ is a paramagnet (PM) above the transition temperature $T_C \sim 177$ K, below which there is an open question on whether the magnetic phase is a ferromagnetic phase or a mixture of ferromagnetic (FM) and antiferromagnetic (AFM) phases. The AFM phase can exist in two possible configurations called "In/Out" and "Chiral" AFM phase. b. Synchrotron based $k_z$ dependent ARPES result showing general agreement with the c-axis ferromagnetic band structure calculation assuming $\mu_{Co} = 0.35\ \mu_B$.

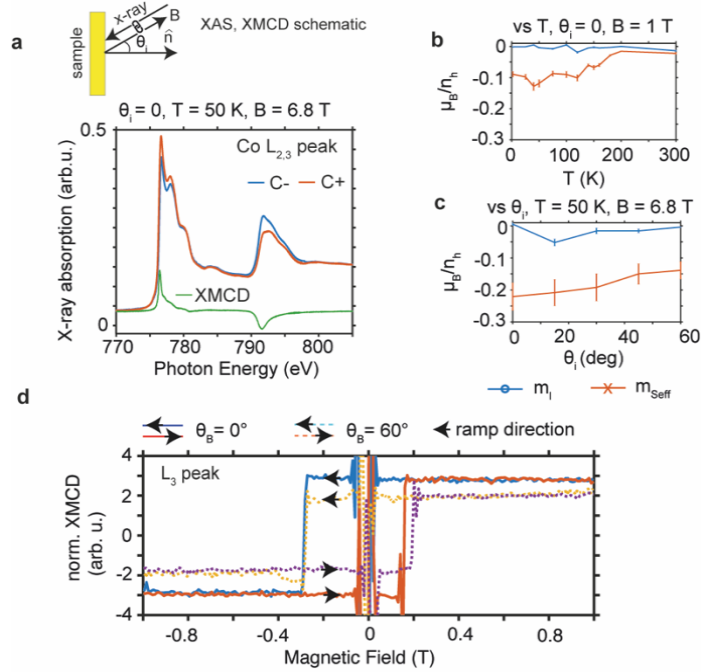

**Figure 2.** (a) Schematic of XAS measurement where the x-ray and the B field are colinear and antiparallel to each other. The XAS result of the Co $L_{2,3}$ peak measured using total electron yield from both circular polarizations show a clear dichroism. (b) Temperature dependent sum-rule analysis of the XMCD Co peak showing that the orbital component is negligible and the magnetism is dominated by the spin-effective component, which vanishes around 175 K in accordance with the magnetic phase transition. (c) Angle dependent sum-rule analysis shows a negligible orbital component while the value of the effective spin drops to $\sim \frac{2}{3}$ at 60° x-ray incident angle relative to normal incidence, suggesting the system has a strong magnetic anisotropy and largely maintains the moment along the c-axis at rotating fields of 6.8 T despite the coercive field for magnetization along the c axis being one order of magnitude smaller. (d) The hysteresis curve at 0° and 60° x-ray incident angles at T = 50 K obtained by tracing the normalized XMCD $L_3$ peak shows the signal saturation and a relatively rectangular shape with a visible exchange bias. The saturated signal shows a reduction to $\sim \frac{2}{3}$ at 60° x-ray incident angle compared to normal incidence indicating that the magnetic moment does not fully follow the magnetic field direction. The exchange bias suggests the existence of an AFM phase.

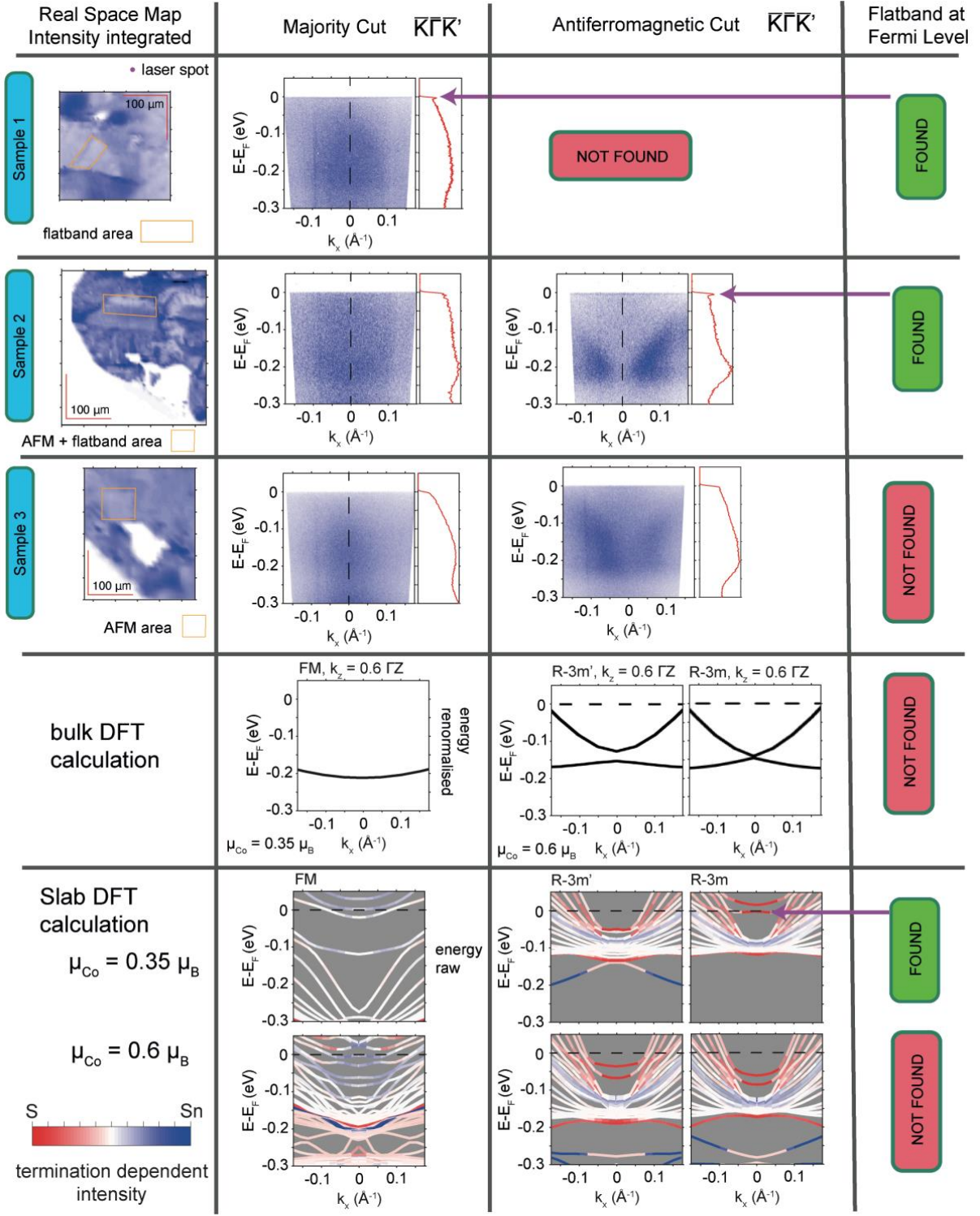

**Figure 3**. Summary of the real-space map obtained from the total integrated intensity, example of the $(E, k)$ dispersion along the $\overline{K}\overline{\Gamma}\overline{K}$ of the ferromagnetic band in the majority of the area, example of $(E, k)$ dispersion along the $\overline{K}\overline{\Gamma}\overline{K}$ of the antiferromagnetic band located at the yellow box of the spatial map, and checklist of flat band at $E_F$ for the three different cleaved samples of $Co_3Sn_2S_2$. The next row is the bulk DFT calculation for the pure ferromagnetic system and the two possible antiferromagnetic phases of $Co_3Sn_2S_2$ at the $k_z$ position probed by the laser energy (6.01 eV) using effective $\mu = 0.35\ \mu_B$ for the FM phase and $\mu = 0.6\ \mu_B$ for the in-plane AFM phase. The last row is the slab calculation with the color-scale indicating the location of the calculated band (from the S termination towards the Sn termination), for both $\mu_{Co} = 0.35\ \mu_B$ and $\mu_{Co} = 0.6\ \mu_B$. The sulphur termination with $\mu_{Co} = 0.35\ \mu_B$ hosts a relatively flat band at $E_F$ which could explain the flat band observed in the ARPES measurement.

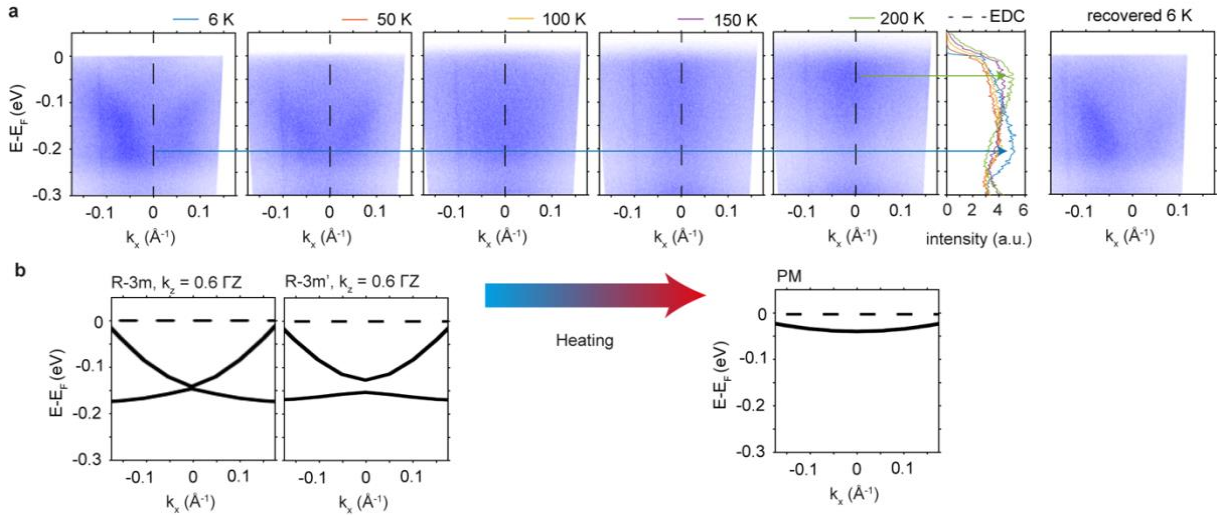

**Figure 4.** a. Temperature dependent dispersion cut in the antiferromagnetic region showing that the butterfly shape transforms into a broad band close to the Fermi energy $E_F$, whose EDC can be traced to shift up closer to $E_F$ upon warming, agreeing with the paramagnetic band position from the bulk DFT calculation (b). This butterfly shape is reproducible around the same area after re-cooling.